



\input lanlmac
\input amssym

\newcount\figno
\figno=0
\def\fig#1#2#3{
\par\begingroup\parindent=0pt\leftskip=1cm\rightskip=1cm\parindent=0pt
\baselineskip=11pt
\global\advance\figno by 1
\midinsert
\epsfxsize=#3
\centerline{\epsfbox{#2}}
\vskip 12pt
\centerline{{\bf Fig. \the\figno:~~} #1}\par
\endinsert\endgroup\par
}
\def\figlabel#1{\xdef#1{\the\figno}}


\def\th{\theta}
\def\Th{\Theta}
\def\ep{\epsilon}

\def\S{{\bf S}}

\def\hf{{1\over 2}}

\def\R{{\bf R}}
\def\o{\over}
\def\til#1{\widetilde{#1}}
\def\si{\sigma}
\def\Si{\Sigma}
\def\b#1{\overline{#1}}
\def\del{\partial}
\def\wg{\wedge}

\def\lf{\left}
\def\ri{\right}
\def\riya{\rightarrow}

\def\ga{\gamma}

\def\al{\alpha}
\def\om{\omega}

\def\dag{\dagger}
\def\rt#1{\sqrt{#1}}

\def\Hom{\rm Hom}

\def\sitarel#1#2{\mathrel{\mathop{\kern0pt #1}\limits_{#2}}}

\def\cob{\delta}

\lref\HarveyAB{
  J.~A.~Harvey and S.~Jensen,
  ``Worldsheet instanton corrections to the Kaluza-Klein monopole,''
  arXiv:hep-th/0507204.
}
\lref\GregoryTE{
  R.~Gregory, J.~A.~Harvey and G.~W.~Moore,
  ``Unwinding strings and T-duality of Kaluza-Klein and H-monopoles,''
  Adv.\ Theor.\ Math.\ Phys.\  {\bf 1}, 283 (1997)
  [arXiv:hep-th/9708086].
}
\lref\TongRQ{
  D.~Tong,
  ``NS5-branes, T-duality and worldsheet instantons,''
  JHEP {\bf 0207}, 013 (2002)
  [arXiv:hep-th/0204186].
}
\lref\GibbonsNT{
  G.~W.~Gibbons and P.~Rychenkova,
  ``HyperKaehler quotient construction of BPS monopole moduli spaces,''
  Commun.\ Math.\ Phys.\  {\bf 186}, 585 (1997)
  [arXiv:hep-th/9608085].
}
\lref\SenZB{
  A.~Sen,
  ``Kaluza-Klein dyons in string theory,''
  Phys.\ Rev.\ Lett.\  {\bf 79}, 1619 (1997)
  [arXiv:hep-th/9705212].
}
\lref\RubackAG{
  P.~J.~Ruback,
  ``The Motion Of Kaluza-Klein Monopoles,''
  Commun.\ Math.\ Phys.\  {\bf 107}, 93 (1986).
}
\lref\HawkingJB{
  S.~W.~Hawking,
  ``Gravitational Instantons,''
  Phys.\ Lett.\ A {\bf 60}, 81 (1977).
}
\lref\GibbonsXM{
  G.~W.~Gibbons and S.~W.~Hawking,
  ``Classification Of Gravitational Instanton Symmetries,''
  Commun.\ Math.\ Phys.\  {\bf 66}, 291 (1979).
}
\lref\RocekPS{
  M.~Rocek and E.~P.~Verlinde,
  ``Duality, quotients, and currents,''
  Nucl.\ Phys.\ B {\bf 373}, 630 (1992)
  [arXiv:hep-th/9110053].
}
\lref\GatesNK{
  S.~J.~.~Gates, C.~M.~Hull and M.~Rocek,
  ``Twisted Multiplets And New Supersymmetric Nonlinear Sigma Models,''
  Nucl.\ Phys.\ B {\bf 248}, 157 (1984).
}
\lref\GibbonsIY{
  G.~W.~Gibbons, G.~Papadopoulos and K.~S.~Stelle,
  ``HKT and OKT geometries on soliton black hole moduli spaces,''
  Nucl.\ Phys.\ B {\bf 508}, 623 (1997)
  [arXiv:hep-th/9706207].
}
\lref\SorkinNS{
  R.~d.~Sorkin,
  ``Kaluza-Klein Monopole,''
  Phys.\ Rev.\ Lett.\  {\bf 51}, 87 (1983).
}
\lref\GrossHB{
  D.~J.~Gross and M.~J.~Perry,
  ``Magnetic Monopoles In Kaluza-Klein Theories,''
  Nucl.\ Phys.\ B {\bf 226}, 29 (1983).
}
\lref\BanksRJ{
  T.~Banks, M.~Dine, H.~Dykstra and W.~Fischler,
  ``Magnetic Monopole Solutions Of String Theory,''
  Phys.\ Lett.\ B {\bf 212}, 45 (1988).
}
\lref\GauntlettNN{
  J.~P.~Gauntlett, J.~A.~Harvey and J.~T.~Liu,
  ``Magnetic monopoles in string theory,''
  Nucl.\ Phys.\ B {\bf 409}, 363 (1993)
  [arXiv:hep-th/9211056].
}
\lref\KhuriWW{
  R.~R.~Khuri,
  ``A Multimonopole solution in string theory,''
  Phys.\ Lett.\ B {\bf 294}, 325 (1992)
  [arXiv:hep-th/9205051].
}
\lref\OoguriWJ{
  H.~Ooguri and C.~Vafa,
  ``Two-Dimensional Black Hole and Singularities of CY Manifolds,''
  Nucl.\ Phys.\ B {\bf 463}, 55 (1996)
  [arXiv:hep-th/9511164].
}
\lref\WittenYC{
  E.~Witten,
  ``Phases of N = 2 theories in two dimensions,''
  Nucl.\ Phys.\ B {\bf 403}, 159 (1993)
  [arXiv:hep-th/9301042].
}
\lref\KapustinHA{
  A.~Kapustin and M.~J.~Strassler,
  ``On mirror symmetry in three dimensional Abelian gauge theories,''
  JHEP {\bf 9904}, 021 (1999)
  [arXiv:hep-th/9902033].
}
\lref\HauselXG{
  T.~Hausel, E.~Hunsicker and R.~Mazzeo,
  ``Hodge cohomology of gravitational instantons,''
  arXiv:math.dg/0207169.
}
\lref\ColemanCI{
  S.~R.~Coleman,
  ``There Are No Goldstone Bosons In Two-Dimensions,''
  Commun.\ Math.\ Phys.\  {\bf 31}, 259 (1973).
}
\lref\CherkisXC{
  S.~A.~Cherkis and A.~Kapustin,
  ``D(k) gravitational instantons and Nahm equations,''
  Adv.\ Theor.\ Math.\ Phys.\  {\bf 2}, 1287 (1999)
  [arXiv:hep-th/9803112];
  ``Singular monopoles and gravitational instantons,''
  Commun.\ Math.\ Phys.\  {\bf 203}, 713 (1999)
  [arXiv:hep-th/9803160].
}
\lref\CallanAT{
  C.~G.~.~Callan, J.~A.~Harvey and A.~Strominger,
  ``Supersymmetric string solitons,''
  arXiv:hep-th/9112030.
}
\lref\SeibergZK{
  N.~Seiberg,
  ``New theories in six dimensions and matrix description of M-theory on  T**5
  and T**5/Z(2),''
  Phys.\ Lett.\ B {\bf 408}, 98 (1997)
  [arXiv:hep-th/9705221].
}
\lref\SeibergTD{
  N.~Seiberg and S.~Sethi,
  ``Comments on Neveu-Schwarz five-branes,''
  Adv.\ Theor.\ Math.\ Phys.\  {\bf 1}, 259 (1998)
  [arXiv:hep-th/9708085].
}
\Title{             
                                             \vbox{
                                             \hbox{hep-th/0508097}}}
{\vbox{
\centerline{Linear Sigma Models of H and KK Monopoles}
}}

\vskip .2in

\centerline{Kazumi Okuyama}

\vskip .2in

\centerline{Department of Physics and Astronomy, 
University of British Columbia} 
\centerline{Vancouver, BC, V6T 1Z1, Canada}
\centerline{\tt kazumi@phas.ubc.ca}
\vskip 3cm
\noindent

\noindent
We propose a gauged linear sigma model of $k$ H-monopoles.
We also consider the T-dual of this model describing
KK-monopoles and clarify the meaning of
``winding coordinate'' studied recently in hep-th/0507204.
\Date{August 2005}

\vfill
\vfill

\newsec{Introduction}
The T-duality between H-monopoles (NS5-branes on $\S^1$) \refs{\BanksRJ,\GauntlettNN,\KhuriWW}
and KK-monopoles (multi-Taub-NUT space) \refs{\SorkinNS,\GrossHB} is a well-established duality \OoguriWJ.
At the level of supergravity this duality exchanges the two gauge fields
$B_{i9}$ and $g_{i9}$ where $i=6,7,8$ and $x^9$ is the $\S^1$ direction 
(we take the worldvolume of monopoles to be extended in the 
$012345$ directions).

However, it was pointed out \GregoryTE\
 that to fully understand the correspondence
of collective coordinates of these monopoles we have to
take into account of the stringy effects.
A single H-monopole carries collective coordinates $\R^3\times\S^1$
which correspond to the transverse space of the NS5-brane 
\foot{We consider
NS5-branes in Type IIB theory. NS5-branes in Type IIA theory have
extra $\S^1$ moduli \refs{\SeibergZK,\SeibergTD}.}.
On the other hand, the Taub-NUT space has a moduli $\R^3$ which specifies
the degeneration locus of the $\S^1$ fiber and hence 
$\S^1$ part of the moduli on the H-monopole side
is missing in the geometric picture of KK-monopoles.
In other words, KK-monopoles are smeared in the $\S^1$ direction and 
there is no definite position in $\S^1$.
It is known that the missing moduli comes 
from the zero-modes of the B-field given by the
harmonic two-forms on the Taub-NUT space \SenZB.
Since B-field couples to the winding number,
the $\S^1$ moduli of KK-monopole is not a geometric translation
mode but the shift symmetry in the ``winding coordinate'' \GregoryTE.

In a recent paper \HarveyAB, it is argued that 
a gauged linear sigma model (GLSM) \WittenYC\ is a useful tool
to analyze this problem.
In \HarveyAB\ the GLSM of a single KK-monopole was studied
by taking the T-dual of the GLSM for a
H-monopole obtained by Tong \TongRQ, and 
the worldsheet instanton corrections to the
effective sigma-model metric is computed.
This result seems puzzling from the effective nonlinear sigma model
perspective since there is no
closed two-cycle in a single KK-monopole background
where the worldsheet is supposed to wrap.

In this paper, we generalize \refs{\TongRQ,\HarveyAB}
and construct the GLSM of multiple H-monopoles and its T-dual GLSM
of the multiple KK-monopoles.
We argue that the moduli of H- and KK-monopoles appear 
in GLSM as parameters of two-dimensional field theory and
clarify the physical meaning of ``winding moduli'' of KK-monopoles
by analyzing the effective theory obtained from the
GLSM. We also suggest a possible resolution to
the above-mentioned puzzle.

The organization of this paper is as follows.
In section 2, we construct a GLSM for $k$ H-monopoles and study its
low energy behavior and instanton corrections.
In section 3, by taking the T-duality of H-monopole GLSM we obtain the
GLSM for $k$ KK-monopoles. In section 4, we discuss the 
instanton effect of KK-monopole model from the effective theory perspective.

\newsec{GLSM for H-monopoles}
In this section, we construct a GLSM for the $k$ H-monopoles 
by generalizing the known model for $k=1$ \TongRQ.
Our model is a ${\cal N}=(4,4)$ $U(1)^k$ gauge theory
with $k$ charged hypermultiplets and one neutral twisted hypermultiplet.

\subsec{The Model}
We use the ${\cal N}=2$ superfield notation.
We argue that $k$ H-monopoles are described by the GLSM
with the following $D$-term ${\cal L}_D$, the superpotential
term ${\cal L}_F$, and the twisted superpotential term ${\cal L}_{\til{F}}$:
\eqn\LHa{\eqalign{
{\cal L}_D&=\int d^4\th {1\o g^2}(-\Th^\dag\Th+\Psi^\dag\Psi)
+\sum_{a=1}^k\lf\{{1\o e_a^2}(-\Si^\dag_a\Si_a+\Phi_a^\dag\Phi_a)
+Q^\dag_ae^{V_a}Q_a+\til{Q}^\dag_ae^{-V_a}\til{Q}_a\ri\}\cr
{\cal L}_F&=\int d\th^+d\th^- \sum_{a=1}^k\Big\{\til{Q}_a\Phi_aQ_a
+(s_a-\Psi)\Phi_a\Big\},\qquad 
{\cal L}_{\til{F}}=\int d\th^+d\bar{\th}^-\sum_{a=1}^k(t_a-\Th)\Si_a
}}
Let us explain our notation in \LHa.
$(\Phi_a,\Si_a)$ is the ${\cal N}=4$ $U(1)$ vector multiplet,
$(Q_a,\til{Q}_a)$ is the charged hypermultiplet, and
$(\Psi,\Th)$ is the neutral twisted hypermultiplet.
We can see that \LHa\ reduces to the model in \TongRQ\ when we set $k=1$. 
$(s_a,t_a)$ in ${\cal L}_F$ and ${\cal L}_{\til{F}}$
are some constant parameters
corresponding to
the FI-parameters and the theta-angle 
\eqn\taparam{
s_a=r^1_a+ir^2_a,\qquad 
t_a=r^3_a+i\th_a.
}
In the following, we use the lower case letters for the scalar field component
of the superfield except for the twisted hypermultiplet.
Following \TongRQ, we denote the scalar 
component of twisted hypermultiplet as $(r^1,r^2,r^3,\th)$:
\eqn\twchi{\eqalign{
\Psi&=r^1+ir^2+\rt{2}\th^+\chi_++\rt{2}\th^-\chi_-+\cdots\cr
\Th&=r^3+i\th-i\rt{2}\th^+\b{\til{\chi}}_+
-i\rt{2}\bar{\th}^-\til{\chi}_-+\cdots.
}}
We also use the vector notation $\vec{r}=(r^1,r^2,r^3)$
to emphasize the triplet of R-symmetry $SU(2)_R$.
The important property of this Lagrangian is that
the scalar fields $(\vec{r},\th)$ act as the dynamical FI-parameters
and the dynamical theta-angle. 

It is easy to write down the component field
expression of the Lagrangian in
the Wess-Zumino gauge.
Here we focus on the bosonic part of the action
\eqn\action{
S={1\o2\pi}\int d^2x({\cal L}_{\rm kin}+{\cal L}_{\rm pot}+{\cal L}_{\rm top}
+{\cal L}_{\rm fermion})
}
where ${\cal L}_{\rm kin}$ is the kinetic term of bosonic fields
\eqn\Lkin{
{\cal L}_{\rm kin}=-{1\o 2g^2}\Big((\del\vec{r})^2+(\del\th)^2\Big)
+\sum_{a=1}^k\lf\{{1\o 2e_a^2}\Big((F_{01}^a)^2-|\del\phi_a|^2-|\del\si_a|^2
\Big)-|{\cal D}q_a|^2-|{\cal D}\til{q}_a|^2\ri\}
}
and ${\cal L}_{\rm pot}$ is the potential term
\eqn\Lpot{\eqalign{
{\cal L}_{\rm pot}=\sum_{a=1}^k\Big\{&-{e_a^2\o2}(|q_a|^2-|\til{q}_a|^2-r^3+r_a^3)^2
-{e_a^2\o2}|2q_a\til{q}_a-(r^1+ir^2)+(r_a^1+ir_a^2)|^2\cr
&-(|\phi_a|^2+|\si_a|^2)(|q_a|^2+|\til{q}_a|^2+g^2)\Big\}
}}
and ${\cal L}_{\rm top}$ is the topological term
\eqn\Ltop{
{\cal L}_{\rm top}=\sum_{a=1}^k(\th-\th_a)F_{01}^a.
}
Note that due to this axionic coupling the field $\th$ becomes a compact
boson $\th\sim\th+2\pi$. Note also that the total space of parameters
$\{(\vec{r}_a,\th_a)\}_{a=1}^k$ is $(\R^3\times \S^1)^k/S_k$. The quotient by
$S_k$ comes from the fact that the permutation of the $k$ gauge groups 
leads to the same theory.
We will see in section 2.3 that this parameter space has the spacetime 
interpretation as the moduli space of $k$ H-monopoles.

\subsec{The Low Energy Theory}
Now let us consider the effective theory of this model
by restricting the Lagrangian to the massless modes.
From the expression of ${\cal L}_{\rm pot}$ \Lpot, the moduli space of 
vacua\foot{Strictly speaking, there is no moduli space of vacua in
 two dimensions because of the Coleman theorem \ColemanCI. 
We analyze the low energy theory in the spirit 
of Born-Oppenheimer approximation.
} is characterized by the equations
\eqn\vaceq{
F_{01}^a=\si_a=\phi_a=0,\quad |q_a|^2-|\til{q}_a|^2=r^3-r_a^3,\quad
2q_a\til{q}_a=r^1+ir^2-(r^1_a+ir_a^2).
}
From the last two equations in \vaceq, it follows that
\eqn\qptilqvac{
|q_a|^2+|\til{q}_a|^2=|\vec{r}-\vec{r}_a|.
}
In the IR limit $e_a\riya\infty$, the vector multiplet 
and charged hypermultiplets become massive and they can be integrated out.
The crucial step is to rewrite the kinetic term of hypermultiplet
restricted on the locus \vaceq
\eqn\Dqsq{
|{\cal D}q_a|^2+|{\cal D}\til{q}_a|^2={1\o4|\vec{r}-\vec{r}_a|}(\del\vec{r})^2
+{|\vec{r}-\vec{r}_a|\o4}(2A^a+2\del\varphi_a+\vec{\om}_a\cdot\del\vec{r})^2
}
where $\varphi_a=-{\rm arg}(iq_a)$ and $\vec{\om}_a$ is given by
\eqn\omarot{
\nabla\times\vec{\om}_a=\nabla{1\o|\vec{r}-\vec{r}_a|}.
}
After setting ${1\o2e_a^2}(F_{01}^a)^2\riya0$, the action becomes
 quadratic in $A_\mu^a$.
Therefore we can integrate out the gauge field classically and
we find
\eqn\AsolH{
A_\mu^a=-\del_\mu\varphi_a
-\hf\vec{\om}_a\cdot\del_\mu\vec{r}+{1\o2|\vec{r}-\vec{r}_a|}
\ep_{\mu\nu}\del^\nu\th.
}
Plugging this back into the original Lagrangian,
the effective Lagrangian for the twisted hypermultiplet is found to be
\eqn\Lkineff{
{\cal L}_{\rm eff}=\hf H(\del_\mu\vec{r}\cdot\del^\mu\vec{r}
+\del_\mu\th\del^\mu\th)
+\ep_{\mu\nu}\vec{\om}\cdot\del^\mu\vec{r}\del^\nu\th
}
where
\eqn\Hom{
H={1\o g^2}+\hf\sum_{a=1}^k{1\o|\vec{r}-\vec{r}_a|},\quad
\vec{\om}=\hf\sum_{a=1}^k\vec{\om}_a,\quad\nabla\times\vec{\om}=\nabla H.
}
This Lagrangian \Lkineff\ is nothing but 
the nonlinear sigma model describing the $k$ NS5-branes smeared in 
the $\S^1_\th$ direction. The last term in \Lkineff\ represents
 the coupling to the B-field generated by the NS5-branes.
As discussed in \TongRQ, the localization in the $\th$ direction
is recovered by summing over the instanton effects, which we will turn next.

\subsec{Worldsheet Instanton Corrections}
The instanton effects in the $k=1$ model is studied in \TongRQ.
In the case of $k$ H-monopoles, we expect that the 
instanton effects in different $U(1)$ sectors
are decoupled from each other
in the first approximation.
Therefore, we can simply use the result of \TongRQ\ for the single $U(1)$ 
theory and sum over the $k$ $U(1)$ contributions. The instanton action
in the $a^{\rm th}$ $U(1)$ sector is given by \TongRQ
\eqn\Sna{
S_{n_a}^{\rm inst}=|n_a||\vec{r}-\vec{r_a}|-in_a(\th-\th_a)
}
where $n_a$ is the instanton number in the $a^{\rm th}$ $U(1)$ gauge field
\eqn\nadef{
n_a=-\int_\Si{F^a\o2\pi}.
}
In general it is quite difficult to compute the $n$-instanton effect
since it involves a complicated integral over the instanton moduli space.
However,  it is conjectured
that all instanton numbers contribute with equal weight
in the effective metric \TongRQ\ to match the target space picture.
Applying this conjecture to our case, we expect that $H$
in \Hom\ is corrected to
\eqn\Hcorrected{\eqalign{
H&={1\o g^2}+\hf\sum_{a=1}^k
{1\o|\vec{r}-\vec{r_a}|} 
\sum_{n_a\in{\Bbb Z}}e^{-|n_a||\vec{r}-\vec{r_a}|+in_a(\th-\th_a)} \cr
&={1\o g^2}+\hf\sum_{a=1}^k{\sinh|\vec{r}-\vec{r_a}|\o|\vec{r}-\vec{r_a}|}
{1\o
\cosh|\vec{r}-\vec{r_a}|-\cos(\th-\th_a)}\cr
 &={1\o g^2}+\sum_{a=1}^k
\sum_{m_a\in{\Bbb Z}}{1\o |\vec{r}-\vec{r_a}|^2+(\th-\th_a+2\pi m_a)^2}.
}}
The sigma model \Lkineff\ with this instanton corrected
$H$ \Hcorrected\ corresponds to the $k$ NS5-branes localized
at $(\vec{r}_a,\th_a)\in \R^3\times\S^1$. 
From the last expression in \Hcorrected, we 
can see that this corresponds to the periodic array of CHS solution \CallanAT.
This effective theory preserves a ${\cal N}=(4,4)$ SUSY since it is
a nonlinear sigma model of hyperK\"{a}hler with torsion 
\refs{\GatesNK,\GibbonsIY}. 
As is clear from the derivation, the collective coordinates
$(\vec{r}_a,\th_a)$ of NS5-branes appear in the GLSM as the FI-parameter and
the theta-parameter.

From the first expression of $H$ in \Hcorrected, we can see that
the instanton number $n_a$ has a spacetime interpretation as
the momentum along $\S^1_\th$ and the collective coordinates
$\th_a$'s are conjugate to this momentum mode.
In other words, $\th_a$ is the translation mode in the $\S^1_\th$ direction.

\newsec{GLSM for KK-monopoles}
In this section, we consider the T-dual of H-monopole GLSM \LHa\
following the general prescription \RocekPS.
We dualize the ${\cal N}=2$ twisted chiral multiplet $\Th$ as was done for
the $k=1$ case \HarveyAB.
To do this, we first rewrite the $F$-term involving $\Th$ as a $D$-term
\eqn\ThVa{
\int d^4\th-{1\o g^2}\Th^\dag\Th-(\Th+\Th^\dag)\sum_{a=1}^kV_a.
}
Then we use the usual trick of introducing a real superfield $B$
and a chiral superfield $P$
\eqn\BVa{
\int d^4\th-{1\o2g^2}B^2-{1\o g^2}\Big(P+P^\dag+g^2\sum_{a=1}^kV_a\Big)B.
}
\BVa\ goes back to \ThVa\ by integrating out $P$, while
integrating out $B$ leads to 
\eqn\BintP{
\int d^4\th{1\o2g^2}\Big(P+P^\dag+g^2\sum_{a=1}^kV_a\Big)^2.
}
Therefore, the T-dual of \LHa\ is given by
\eqn\LagsKK{\eqalign{
{\cal L}_D&=\int d^4\th {1\o g^2}\Psi^\dag\Psi+
{1\o2g^2}\Big(P+P^\dag+g^2\sum_{a=1}^kV_a\Big)^2 \cr
&\qquad+\sum_{a=1}^k\lf\{{1\o e_a^2}(-\Si^\dag_a\Si_a+\Phi_a^\dag\Phi_a)
+Q^\dag_ae^{V_a}Q_a+\til{Q}^\dag_ae^{-V_a}\til{Q}_a\ri\}\cr
{\cal L}_F&=\int d\th^+d\th^- \sum_{a=1}^k\Big\{\til{Q}_a\Phi_aQ_a
+(s_a-\Psi)\Phi_a\Big\},\qquad 
{\cal L}_{\til{F}}=\int d\th^+d\bar{\th}^-\sum_{a=1}^kt_a\Si_a.
}}
Again it is easy to write down the Lagrangian in the component fields.
The bosonic part of the Lagrangian reads
\eqn\LkinpotKK{\eqalign{
{\cal L}_{\rm kin}&=
-{1\o 2g^2}(\del\vec{r})^2+{g^2\o2}\Big(\del\ga+\sum_{a=1}^kA^a\Big)^2\cr
&+\sum_{a=1}^k\lf\{{1\o 2e_a^2}\Big((F_{01}^a)^2-|\del\phi_a|^2-|\del\si_a|^2
\Big)-|{\cal D}q_a|^2-|{\cal D}\til{q}_a|^2\ri\} \cr
{\cal L}_{\rm pot}&=\sum_{a=1}^k\Big\{-{e_a^2\o2}(|q_a|^2-|\til{q}_a|^2-r^3+r_a^3)^2
-{e_a^2\o2}|2q_a\til{q}_a-(r^1+ir^2)+(r_a^1+ir_a^2)|^2\cr
&\qquad-(|\phi_a|^2+|\si_a|^2)(|q_a|^2+|\til{q}_a|^2+g^2)\Big\} \cr
{\cal L}_{\rm top}&=-\sum_{a=1}^k\th_aF_{01}^a,
}}
where we denote the scalar component of $P$ as
$r^3+ig^2\ga$.

Let us comment on some of the properties of this Lagrangian.
One important property of this model
is that the scalar field $\ga$ is shifted by all $k$
$U(1)$ gauge transformations
\eqn\gaugega{
\cob A^a_\mu=\del_\mu\al_a,\quad\cob\ga=-\al_a \quad(a=1,\cdots,k).
}
Combining $\ga$ with the phase $\varphi_a=-\arg(iq_a)$ of 
the hypermultiplet scalar, we can define
the gauge invariant field $\kappa$ as
\eqn\kainv{
\kappa=\ga-\sum_{a=1}^k\varphi_a.
}
Note that there is no dynamical theta angle $\th$ 
in ${\cal L}_{\rm top}$ \LkinpotKK\ since we have dualized 
the twisted chiral multiplet $\Th$. It is also worth mentioning that
the potential term ${\cal L}_{\rm pot}$ is exactly the same as
the H-monopole model \Lpot.
Therefore, the equation for the moduli space of vacua is the same
as the H-monopole case \vaceq.

By looking at the gauge symmetry of this model \LagsKK, 
we immediately notice that this system gives the hyperK\"{a}hler quotient
construction of multi-Taub-NUT space \GibbonsNT
\foot{A similar model in three dimensions was considered
 in \KapustinHA.}.
We can explicitly write down the low energy theory following the
procedure in the previous section.
By restricting to the vacuum \vaceq\ and integrating out the gauge fields,
we find that $A^a$ is given by
\eqn\Aaout{
A_\mu^a=-{1\o 2|\vec{r}-\vec{r}_a|H}(\del_\mu\kappa
-\vec{\om}\cdot\del_\mu \vec{r})-\del_\mu\varphi_a
-\hf\vec{\om}_a\cdot\del_\mu\vec{r}.
}
where $\vec{\om}_a$ and $H,\vec{\om}$ are defined in
\omarot\ and \Hom.
Plugging this back into the Lagrangian,
the low energy effective Lagrangian is given by 
\eqn\LeffKK{
{\cal L}_{\rm eff}=\hf H(\del\vec{r})^2+{1\o 2H}
(\del\kappa-\vec{\om}\cdot\del\vec{r})^2.
}
This is nothing but the  metric of the
multi-Taub-NUT space $M_k$ 
\refs{\HawkingJB,\GibbonsXM}.
As expected from the supergravity analysis of the B-field zero-mode \SenZB,
the curvature $F^a=dA^a$ of the forms \Aaout\ are 
equal to the $k$ linearly independent $L^2$-normalizable self-dual 
harmonic two-forms
on the multi-Taub-NUT space \RubackAG \foot{We make slight abuse
of notation that we identify the forms on the effective target space 
(i.e. Taub-NUT) and their pull-back to the worldsheet.}
\foot{
Note the difference of the $L^2$ harmonic two-forms on
the ALF space (Taub-NUT) and the ALE space \HauselXG
$$
{\rm dim}L^2{\cal H}^2({\rm ALF}_{A_{k-1}})=k,\quad
 {\rm dim}L^2{\cal H}^2({\rm ALE}_{A_{k-1}})=k-1.
$$
}.

\newsec{The Physical Meaning of the Collective Coordinate $\th_a$}
We have seen in section 2 that the instanton numbers
in the H-monopole model correspond to
the momentum modes in the target space and the collective coordinate
$\th_a$ (or theta-parameter in GLSM) is the translation mode in the
$\S^1_\th$. From the general property of T-duality, we expect
that the instanton number in the KK-monopole model corresponds
to the winding number in the target space picture
and the conjugate collective coordinate $\th_a$ is the translation
mode in the ``winding coordinate'' $\S^1_{\tilde{\th}}$.\foot{
We do not try to define $\til{\th}$ as a worldsheet {\it field} \HarveyAB. 
Rather, the collective coordinates $\th_a$ appear in the worldsheet action
as {\it parameters}. By construction, the GLSM of H-monopoles and the GLSM of
KK-monopoles have the same parameter space $(\R^3\times\S^1)^k/S_k$.}
The instanton corrections in the KK-monopole model is analyzed in
the UV gauge theory language in \HarveyAB\ for the $k=1$ case.
In this section we consider the instanton effect from the low energy
sigma model perspective.

As mentioned in the introduction, from the effective theory
point of view it seems puzzling to have a
nontrivial instanton correction for the $k=1$ case
since there is no closed two-cycle in that case.
Here we propose that the relevant instanton correction is coming from
a disk instanton. The possibility of disk instanton is suggested
in \GregoryTE\ in relation to the description of KK-monopoles in 
the string field theory.

Since we are considering the sigma model on the sphere $\Si=\S^2$,
it seems impossible to have a disk instanton. However,
because of the non-compactness of the target space
we can consider a map which effectively looks like a disk.
We decompose the worldsheet sphere into a disk and one point
\eqn\Sdec{
\S^2=D^2\cup\{\infty\}
}
and consider a map which sends the disk into a finite region of target space
and the $\infty$ of worldsheet to the infinity of target space.
With this understanding, we can talk about the disk instanton.

Let us consider the configuration of
disk instanton which wraps $n$ times on
the semi-infinite cigar in the
target space
\eqn\cylsia{
D^2=\{z;|z|<1\}\longrightarrow
nC_a=\{\vec{r}(z)=\vec{r}_a+f(|z|)\vec{v},\,\kappa(z)=n\,\arg(z)\}
}
where $\vec{v}$ is a constant unit vector specifying the direction
of cigar. The function $f(|z|)$ 
satisfies the boundary condition
\eqn\asyfn{
f(0)=0,\quad f(1)=\infty.
}
$f(|z|)$ can be determined
by minimizing the action but we don't need its explicit form.
We can see that $C_a$ has a shape of cigar by noting that
$\S^1$ fiber is degenerate at $\vec{r}=\vec{r}_a$ and the
radius of $\S^1$ fiber approaches a constant value $g$ at infinity.
If we regard $\tau=1/|z|$ as the Euclidean time coordinate on the worldsheet,
the configuration \cylsia\ represents a process that a string winding 
$n$ times around the $\S^1$ fiber at infinity propagates towards
one of the monopole cores $\vec{r}=\vec{r}_a$ and unwinds at $\tau=\infty$
\GregoryTE.

One can compute the flux of the harmonic two-form
$F^a=dA^a$ \Aaout\ through the cigar $C_a$
\eqn\fluxg{
\int_{nC_a}{F^b\o2\pi}=n\cob_{a,b}.
}
In this computation only the first term in \Aaout\ contributes.
Other terms in \Aaout\ couple to the ``unwinding'' string away from the core
of monopoles \GregoryTE.
We can understand the diagonal property \fluxg\ in a more abstract way.
It is known that the intersection form
of the harmonic two-forms $F^a=dA^a$ given by \Aaout\ is diagonal \RubackAG
\eqn\FaFbcob{
\int_{M_k}{F^a\o2\pi}\wg {F^b\o2\pi}=\cob_{a,b}.
}
Therefor we can define $C_a$ as a Poincare dual of the $F^a$
\eqn\PDual{
C_a=PD\lf(\lf[{F^a\o2\pi}\ri]\ri).
}

From this discussion it is clear that the instanton number
in the low energy picture
is the winding number of disk instanton wrapped around the cigar.
The instanton action in the UV gauge theory picture is obtained in \HarveyAB
\eqn\diskinstS{
S^{\rm inst}=|n||\vec{r}-\vec{r}_a|-in\th_a.
}
We do not try to reproduce this instanton action from the
low energy point of view. However, we note that the relevant instanton
in the UV computation is the constrained instanton 
defined in the $g\riya0$ limit \refs{\TongRQ,\HarveyAB}. 
In this limit the asymptotic radius $g$
of the $\S^1$ fiber goes to zero. Therefore it is possible
that the instaton action calculated in the low energy theory
is finite despite the fact that
the cigar $C_a$ is non-compact.
In general  the low energy theory does not necessarily give the 
same instanton action as the UV theory. However we expect that 
the topological charge should have the same interpretation
in the UV and IR. What we have shown is that the disk instanton wrapped
around $C_a$ has the same topological charge as the gauge theory instanton
of $a^{\rm th}$ $U(1)$.

When $k>1$ one can also consider a sphere instanton
which wraps the closed two-cycle in the Taub-NUT
\eqn\Stwoinst{
\Si=\S^2\longrightarrow C_a-C_b.
}
When the two KK-monopoles coincide $\vec{r}_a=\vec{r}_b$, 
the corresponding cycle $C_a-C_b$ vanishes. From the duality
to H-monopole, we expect that the coincident KK-monopoles would show
the throat behavior. 
It would be interesting to see this explicitly using the GLSM.
It would be also interesting to determine 
the precise form of the instanton correction to the
KK-monopole sigma model. It is observed \HarveyAB\ that 
not only the metric is corrected but also the torsion is generated.
Parhaps it might be useful to consider the topological A-twist
of the KK-monopole GLSM.
It would be also interesting to 
construct a GLSM of the $D_k$-type Taub-NUT space \CherkisXC\
and a ${\cal N}=(4,0)$ GLSM describing a bound state of H-monopoles 
and KK-monopoles.

\vskip10mm
\noindent{\bf Acknowledgment:}
I would like to thank Kazuyuki Furuuchi for encouragement.
I would also like to thank Steuard Jensen and David Tong 
for correspondence.
\listrefs
\bye